# Public emotions on Internet: In case of AIGC

Qinglan Wei [a], Jiayi Li [b], Yuan Zhang [c,*]

[a] School of Data Science and Intelligent Media, Communication University of China, Beijing 100024, China
[b] School of Information and Communication Engineering, , Communication University of China, Beijing 100024, China
[c] State Key Laboratory of Media Convergence and Communication, Communication University of China, Beijing 100024, China

ARTICLE INFO

ABSTRACT

The proliferation of interactive AI like ChatGPT has fueled intense public discourse surrounding AI-generated content (AIGC). While some fear job displacement, others anticipate productivity gains. Social media provides a rich source of data reflecting public opinion, attitudes, and behaviors. By examining the factors influencing collective sentiment toward AIGC on various platforms, we can refine products, marketing, and AI models themselves. Our research utilized a novel system for real-time tracking and detailed visualization of public mood related to AIGC. This system enabled analysis of the dynamics shaping opinions on nine AIGC products across China's three leading social media sites. Our findings reveal a negative correlation between user demographics (age and education) and positive sentiment towards AIGC on Douyin, contrasting with Weibo's susceptibility to the rapid spread of extreme viewpoints. This work uniquely connects group dynamics theory with social media sentiment, offering valuable guidance for managing online opinion and tailoring targeted campaigns.

## 1. Introduction

The rapid advancement of deep learning, reflected in a surge of research publications, has propelled AI-generated content (AIGC) into the global spotlight. Tools like ChatGPT, demonstrating remarkable natural language processing capabilities, have captivated users and sparked both excitement and apprehension. While AIGC offers transformative potential across numerous applications, concerns persist regarding ethical implications, the spread of misinformation, and the potential displacement of human workers. The intense public discourse surrounding AIGC, encompassing diverse opinions and anxieties, necessitates further investigation into its societal impact and the evolving relationship between humans and artificial intelligence.

Emotions significantly influence and shape public perception of emerging AIGC, acting as a key catalyst in the evolution of public opinion (Fu et al., 2020). Social media, as a crucial medium for expressing public sentiment and disseminating information, reveals genuine emotional tendencies towards novel issues, making it an indispensable tool for understanding public opinion. Psychological research demonstrates the powerful influence of groups on individuals, including altering attitudes, beliefs, and values, and potentially even inciting harm (e.g., groupthink). Individuals are more susceptible to influence when a group exhibits a tendency to agree on a particular viewpoint (Marques & Paez, 1994; Marques et al., 2001). In the current social media landscape, recommendation algorithms reinforce shared opinions within user groups, leading individuals to increasingly encounter like-minded perspectives and ultimately contributing to group polarization (Barberá et al., 2015; Cinelli et al., 2021). Public opinion frequently exhibits divergence, particularly on contentious issues such as job displacement and economic downturn resulting from the implementation of emerging technologies in production. Therefore, studying the emotional state surrounding AIGC technology from a group dynamics perspective is crucial.

Basically, "group emotion" means when a bunch of people online all feel the same way about something – it's like a mood that spreads. Most studies so far have looked at groups meeting in person. The big problem now is that online groups are different depending on the social media site – it's all about how each site is set up and how people interact there. We don't really understand why the mood on, say, Twitter, is so different from the mood on Facebook. This makes it hard to figure out what's really going on when we see certain things spreading online.

* Corresponding author.
*E-mail address:* yzhang@cuc.edu.cn (Yuan Zhang).

On the other hand, there's a lot of room for improvement in making the way we measure group emotion more scientific and precise. To be specific, most researchers use traditional methods of calculating sentiment in text to analyze and measure group emotion in real-world social media events – not simulations. This mainly focuses on analyzing and classifying the content of comments, and it doesn't really pay attention to how those comments relate to each other. This creates some big challenges. First, how people feel online depends on how they interact with each other. Looking only at individual accounts gives an incomplete picture. Second, using average or total sentiment scores, which is common, makes the results easily skewed by extreme opinions. Third, current methods struggle to accurately identify group emotion when comments are vague or have mixed feelings.

We're at the cutting edge of developing technology to measure group emotion around trending topics on social media. This is a fundamental problem that needs improvement in group emotion communication analysis. Building on existing research, this paper defines, in a general way, the concepts of macro and micro clusters in social networks. We also introduce a novel system for calculating and visualizing group emotion based on multi-level communication chains. This system can capture data in real-time and show a highly detailed picture of current group emotion. To make it easier to understand how things spread across different platforms, we use our own group emotion calculation model to deeply analyze the similarities and differences in what drives group emotion on various platforms. For trending topics related to AIGC (AI-generated content), we chose three of China's most popular social media platforms – Weibo, Douyin, and Bilibili – and specifically analyzed the group dynamics driving different emotions toward nine AIGC products. By summarizing the spread of group emotion based on platform-specific differences in emotional drivers and investigating potential influencing factors, our research concludes that user age and education level negatively impact group emotion towards AIGC technology on Douyin. Additionally, because of Weibo's unique platform mechanics, extreme emotions are more likely to spread.

## 2. Literature review

### *2.1. Group emotion as a key indication for reflecting social opinions to hot spots*

In the context of Internet group communication, emotional expression possesses a faster and more extensive dissemination capacity, which is inseparable from the internal structure of group communication (Ferrara & Yang, 2015; Cinelli et al., 2021). The concept of group emotion was first proposed by Barsade et al. in 1998. It refers to the consistent emotions exhibited by most members after being communicated and infected by group members (Barsade & Gibson, 1998). In 2006, Maitner et al. proposed the group emotion theory, which considers the interactions between individuals within the group and explains the nature of emotions based on the group and group identity (Maitner et al., 2006; Rydell, 2008). Group emotions include positive emotions, negative emotions, and neutral emotions (Feng et al., 2021).

At present, the research on group emotion in social network mainly focuses on three aspects: the expression of group emotion, the infection of group emotion and the spread of group emotion. The purpose of studying group emotional expression is to explore the internal driving mechanism of emotional expression. This understanding helps manage the evolution of public sentiment by tracking the progress of emotional expression. Dhall initiated research in this area and proposed a group expression model (GEM) for predicting group sentiment (dhall&goecke, 2015; dhall et al., 2015). At present, there are two main methods: bottom-up method, top-down method to calculate overall emotion from individual emotion, and method to calculate group emotion from environmental and group information. These methods, however, have limitations. In terms of quantitative techniques related to group emotions, the current research is limited to real space, mainly relying on deep learning methods. With the rise of deep learning algorithms, neural networks are more and more used in group emotion recognition and classification. For example, Pinto combines facial feature extraction with RESNET to propose a neural network model for extracting group emotions from multiperson videos (Pinto et al. 2020). Sun integrates multimodal feature training models for movies and videos based on LSTM and gem networks, and has achieved promising results on multiple data sets (Sun et al. 2016). Emotional contagion refers to the process by which individuals produce similar emotions after exposure to other people's emotional expressions. The phenomenon is particularly common on social platforms. Kramer and his colleagues confirmed that emotions can spread and infect others through the Internet (Kramer et al. 2014). In the crowd: in a study of popular psychology (1896), Gustav Le Pen asserted that the reciprocal transmission of group emotions determines the choice of group behavior. Instinctive emotions are particularly susceptible to infection, while rational and calm emotions have no effect in groups. Van kleef confirmed the impact of social environment on group emotions and pointed out that the expression and experience of individual emotions are affected by others in the group (van kleef, 2010). Feng believes that emotional contagion is based on a collective culture and values (von scheve&ismer, 2013). Hill et al. Applied the infectious disease transmission model to the emotional transmission process and proposed the Sisa model (Hill&Rand, 2010). On this basis, Liu introduced Sosa SPAA, an enhanced model with broader emotional classification (Liu, Zhang Helan, 2014). Emotional transmission is the process of transforming individual emotions into group emotions, including the mechanism of emotional infection. Group emotional transmission focuses on how individual emotions are transmitted to groups in cyberspace, rather than treating Internet users as a group. Zafarani, for example, pioneered an experiment in emotional communication on the Internet and conducted emotional analysis by obtaining online data from blogs (zafarani, Cole and Liu, 2010). BAC and Lee began studying social platforms and extracting the emotional polarity of each tweet on twitter, the world's largest social networking site, by comparing it with the emotional dictionary to determine the emotional impact of well-known Twitter users (BAE&Lee, 2012). Cai et al. An image social network was constructed on social groups in the network, group emotions were calculated from images posted by users, and the effects of opinion leaders and affinity density on community emotions were simulated (CAI et al. 2018).

It can be seen that the current academic research on group emotional expression, infection and communication mechanism on social networking platforms has a solid theoretical basis. Research shows that group emotions directly affect the formation and evolution of public emotions. However, when analyzing real social network events outside the simulation experiment, most scholars use the traditional text emotion tendency calculation method from the perspective of group emotion analysis and quantification. These methods mainly focus on the emotional analysis and classification of comments, and rarely consider the hierarchical relationship and internal relationship between comments. This method brings several main problems: first, the generation and spread of emotions on social platforms depend on user behavior and interaction. It is one-sided to focus on the emotional expression of the single level account level. Secondly, the current quantitative methods, such as average or cumulative emotional orientation, are deeply affected by extreme emotional reviews. Third, for user reviews with emotional polarity in the gray area, the current quantitative method is easy to lead to inaccurate group emotion recognition in the actual public opinion monitoring.

The purpose of this paper is to tackle the fundamental challenge of quantifying group sentiment in social network event trends, a crucial aspect that requires optimization in the analysis of group emotion communication. Drawing on the extensive theoretical foundation of group emotion in existing research, we aim to establish a multi-level emotion communication chain. Utilizing text emotion analysis technology and user account behavior information, we will delve deeper into the relationship between group sentiment and public emotion. By quantifying the value of group sentiment, we strive to offer fresh perspectives for monitoring and guiding public opinion.

### *2.2. Group dynamics in social networks*

In terms of theoretical research, group dynamics is an interdisciplinary field intersecting with anthropology, political science, pedagogy and sociology. This study mainly focuses on the factors or drivers affecting group behavior, such as group atmosphere, the relationship between group members, leadership style and so on. This paper discusses the nature, characteristics, development law of groups and the relationship between groups, individuals and other groups. For example, Tetlock designed an experiment to verify group dynamics that affect group political decision-making (Tetlock et al. 1992). Abrams conducted subjective group dynamics experiments on group rules in two settings: banks and schools (Abrams et al. 2002). The study of group dynamics based on cyberspace focuses on political and motivational themes. Ernst compared the spread of populism between social media and traditional television media (Ernst et al. 2019), while wahlstrom studied the spread of radical political views on social media (wahlstrom&t2021). Lewis set up a social network on Facebook for a group of college students, using data analysis to study dynamics that affect how people form social relationships (Lewis, Gonzalez and Kaufman, 2012). Zhao Jian proposed a dynamic model of group opinion decision-making in the context of real life and Society (Zhao Jian et al., 2021). At present, the research results of group dynamics on network use motivation are rich, which provide an important reference for our work. Han Jun suggested that groups engaged in online interaction have stronger information consumption motivation, and information interaction will increase their enthusiasm for websites and consumption (KO, CHO&Roberts, 2005). Whiting identified the motivations of the top ten social media users through social surveys (whiting&Williams, 2013); KY Lin's research shows that gender factors affect motivation to use social media (Lin and Lu, 2011). Y Kim acknowledges that users from different countries are affected by cultural environments and have different potential motivations for using the Internet (Kim et al. 2011). Another important aspect of current research on communities formed in cyberspace is the study of communication and universality. Sabat, for example, examined factors affecting the popularity of product content on Facebook, focusing on the impact of posting content through images, videos or product links (Sabat et al. 2014). Welbourne analyzed the drivers of the popularity of educational science videos on youtube, including the number of followers, the length of the video and whether it was recommended by the YouTube recommendation system (welbourne&grant, 2016). Amarasekara conducted a similar study of youtube videos to investigate the impact of moderator or gender on emotional arousal (amarasekara&grant, 2019).

However, most of these studies are confined to a single platform, featuring a limited variety of research content. Previous studies have primarily focused on dynamic indicators and factors related to the motivation for using social networks and the factors influencing the popularity of content spread across the Internet. These studies seldom delve into the causes and driving forces behind the formation of specific group emotions. Furthermore, research on group dynamics within social networks has predominantly been conducted on a single platform, lacking cross-platform comparisons. This limitation hinders the exploration of dynamic differences among various online groups.

Therefore, this paper aims to tackle the problem of comparing group dynamic indicators across multiple platforms to figure out what's *really* driving group emotion. On one hand, we should look at past research on what makes online content popular and create measures to capture the factors shaping group emotion. That's because popularity is a key part of how we measure group emotion. On the other hand, we should broaden our study to include several well-known social media platforms and see how these measures perform on different sites. By doing this, we can explain our findings and understand the dynamics of group emotion.

## 3. Methods

In this section, we describe our proposed method. First, a dataset is built by crawling data from multiple websites that discuss various technical topics related to AIGC. Then a group sentiment computing system is established based on secondary communication. Finally, we select dynamic indicators of different group sentiment formation from the group dynamics theory.

### *3.1. Data collection*

Based on apppc (http://apppc.com/), a data station that evaluates website visits, we select three mainstream and representative platforms for experimentation: Sina Weibo, Bilibili, and Douyin. Sina Weibo is the largest microblog social platform in China, Douyin is the world's largest short video platform widely recognized and applied internationally (Zeng & Abidin, Schäfer, 2021), and Bilibili is the head platform for aggregating videos, representing the forefront of popular fashion culture among young people. All these platforms are characterized by expressions of user-generated media, where users create and share their own content for other users to consume through browsing, liking, commenting, etc (Shao, 2009). The various consumption behaviors of users are the main elements of our group emotional system, and the specific building elements are also different due to the different design of each website system.

When crawling data through keyword search results in these websites, we roughly divide the AIGC technology that needs to be studied into two categories: popular functions and popular products, totaling nine keywords. Popular functions include AI painting, AI writing, AI music, and AI translation. Popular products include Wen Xin Yiyan, New Being, ChatGPT, Dreamilyapp, and Copilot. Among them, AI music refers to the composition and arrangement of music using AI technology, including some imitation vocal singing works after neural network training. AI writing includes AI literary creation, such as poetry and novels, as well as some basic copywriting in people's daily life and work. Microsoft's Copilot provides users a digital companion, offering assistance in generating documents, emails, presentations, and more services in Microsoft 365 softwares. We take comprehensively consideration of selecting keywords to find more useful information. For example, search results for the keyword "AI sequel" are not as comprehensive as those for “Dreamilyapp”. We hypothesize that the group emotion values are different among these topics. Such differences can provide a new view to the further study of group dynamics.

We use a self-developed multi-modal data crawling software on multi-social media platforms, based on Selenium automation and packet capture technology. It can automatically crawl short video or post information and user behavior data under a topic on different platforms. We then perform data preprocessing, which aims to screen and clean of some garbled characters and individual non-related posts in the original download data. We use the Python natural language analysis library SnowNLP to calculate the tendency value through the Bayesian model for all comment texts one by one, and the value range is [-1, 1].

### 3.2. Group emotion calculation

Gestalt psychology is one of the major schools of modern Western psychology, which emphasizes the holistic nature of experience and behavior. It posits that the whole is not equal to the sum of its parts, and the nature of a thing is not determined by any one part but depends on the whole. Moreover, the whole exists independently of its parts (Koffka, 1935). Based on the Gestalt holistic theory, this section aims to provide a general quantitative analysis of the spread of group emotions for hot events on social media platforms. In comparison to the average or cumulative emotion propensity value based on single-level comments, the key to the quantitative method in this paper is to scientifically construct a multi-level emotion propagation chain triggered by the accounts and their posting or commenting behaviors. Based on this emotion communication chain, we calculate the group emotion triggered by posts using the social behavior information of each account user.

#### 3.2.1. Construction of emotional communication Chain

Once a post is published, other users on the platform will like, retweet, and comment on the content of the post, which is regarded as first-level feedback. The user behavior of liking, retweeting, and commenting on first-level comments is regarded as second-level feedback. In this multi-level interaction mechanism, each account, including post publishers, first-level feedback accounts, and second-level feedback accounts, acts as a communicator and conveys its own opinions and emotions. These accounts, whether of official media organizations, weblebrities, or grassroots individuals, can act as passive recipients and active communicators at the same time, which is a characteristic of accounts on social media platforms in the era of group communication.

Based on the above basic concepts of post multi-level emotion transmission chains and social network clusters ( Himelboim et al., 2020), this paper innovatively defines post macro-clusters and micro-clusters. In the macro-cluster, posts and first-level comment accounts serve as the nodes of this heterogeneous network, and the directed edges between the connected nodes serve to characterize the first-level feedback against the content of the post. In the micro-cluster, the first-level comment accounts and second-level comment accounts act as nodes of this homogeneous network, and the directed edges between the connected nodes characterize the second-level feedback for the first-level comments. Because accounts generate emotional infection during the interaction process, local group emotions will be formed in both macro- and micro-clusters.

#### 3.2.2. Cluster emotion computing model

*(1)Cluster Density*

In tightly connected clusters, information spreads and emotions become infected more quickly (Centola, 2010). Therefore, this study describes the closeness of the association between posts and first-level feedback and between first- and second-level feedback by calculating the density values of macro-clusters and micro-clusters to represent the intensity of feedback triggered by posts and first-level comments, respectively. Cluster density is an important metric for describing cluster emotion, denoted as $D$. For the post nodes in the macro-cluster and the first-level comment nodes in the micro-cluster, their entry degrees were calculated by the sum of the number of first-level comments and the number of first-level likes, and the sum of the number of second-level comments and the number of second-level likes, denoted as $Ind$. The quantified formula for the cluster density is: $D = \frac{Ind}{Max(Ind)}$.

It is worth noting that here we use the operation of dividing by $Max(Ind)$ to achieve a larger local emotion calculation weight for clusters with more feedback interactions, and conversely, give a smaller weight to clusters with fewer feedback interactions. This design concept is based on practical analysis. For example, some posts have less first-level feedback (the density of macro-clusters is relatively small); therefore, even if a post is intended to trigger positive social emotions in major events, the post content is not designed properly and does not have a good dissemination effect, resulting in the failure to achieve positive opinion guidance. For example, some first-level comments have almost no second-level feedback (the density of micro-clusters is extremely small), so even if the first-level comment has a highly negative emotional tendency, the comment will be drowned among many comments and will not bring about significant malignant polarized emotion dissemination and infection.

*(2)Cluster Trust*

On the one hand, in an actual social network, there may be machine accounts or water army accounts. On the other hand, some real accounts have relatively random or no clear emotional tendency in posting, commenting, and other behaviors. For these two types of situations, the actual reflection in the post cluster is such that even if there are connected edges between nodes, these edges are either ineffective or inefficient, and the trust degree of the cluster is low overall. Contrariwise, the higher the trust degree, even though the content feedback volume is low, the more likely the content is to trigger polarized group emotion and effectively guide the direction of public opinion. Therefore, cluster trust is another important indicator for describing cluster emotions, denoted as $T$.

Based on the general principles of social network trustworthiness model construction, this study used three indicators, cohesion $T_1$, authority $T_2$, and influence $T_3$, to quantify the cluster trustworthiness of posts or commentary content in the process of emotion transmission. In the calculation process, cohesion depends on the intra-class standard deviation of emotion tendencies among comment texts, and a smaller value, i.e., a higher intra-class similarity, means that the cluster is more cohesive and trustworthy (Lei et al., 2021). The authority depends on the intra-class standard deviation of emotion tendency among comment texts, which characterizes the degree of bifurcation of comment emotion tendency in the cluster, and the higher the degree of bifurcation, reflecting the more controversial the post or comment content is, the less authoritative it is. In the actual quantification process of authority, given that if the emotion of a comment is extremely different from the emotion of its cluster, it may be potentially malicious, the comment is regarded as an outlier and not included in the authority calculation process (Cota et al., 2019). Since the more followers or retweets of the content of the central node account of the cluster, the more likely the expressed content will be accepted by other accounts and the more influential the triggered emotion spread, the influence indicator depends on the normalized number of followers of the account $T_3^{(1)}$ (Zhang and Han, 2021) and the number of retweets of the content $T_3^{(2)}$ (Brin and Page, 1998). The quantitative formula of cluster trust is $T = A_1 T_1 + A_2 T_2 + A_3^{(1)} T_3^{(1)} + A_3^{(2)} T_3^{(2)}$, where $A_1$, $A_2$, $A_3^{(1)}$, and $A_3^{(2)}$ refer to the weights of the number of account fans and the number of retweets for each index of cluster trust cohesiveness, authority, and influence, respectively. In the process of real case analysis, considering the false fans, it is recommended setting the weight $A_3^{(1)}$ of the number of account fans to a lower value when calculating the cluster trust degree.

*(3) Cluster Emotion*

The power of emotion transmission in clusters derives from the emotional resonance triggered by emotional infection, which originates from the flow of information about a cluster's emotional meta. In this study, by adopting the comment text emotion propensity to reflect the emotional meta in the cluster, the average emotion propensity value of the cluster was denoted as $E$, characterizing the degree value of both positive and negative emotion polarity. The cluster average emotion tendency is the third most important metric to describe the cluster emotion.

Cluster emotion is a product of the interaction between different accounts of posts, not just a mechanical aggregation of isolated individual expressions, but also in the process of primary dissemination of posts and secondary dissemination of comment content. The density and trust of the clusters comprising the communication chain determine whether the infection between emotions resonates; for example, if a comment triggers a very negative secondary comment but the trust or cluster density of that micro-cluster is low, that fails to trigger even a small range of malignant polarized emotions, so there is no need to actually make a warning in the public opinion monitoring application. Therefore, this study integrates cluster density, cluster trust, and cluster average emotion tendency and characterizes cluster emotion by $A*T*E$.

Based on the above principles and calculations, the local emotion values of macro- and micro-clusters can be obtained, and the global group emotion triggered by posts can be evaluated comprehensively by the weighting method. The entire process of post group emotion quantification is shown in Figure x. The method proposed in this paper features low computational complexity, is unsupervised, and can easily be migrated to multiple social media platforms, other events, or audiovisual programs.

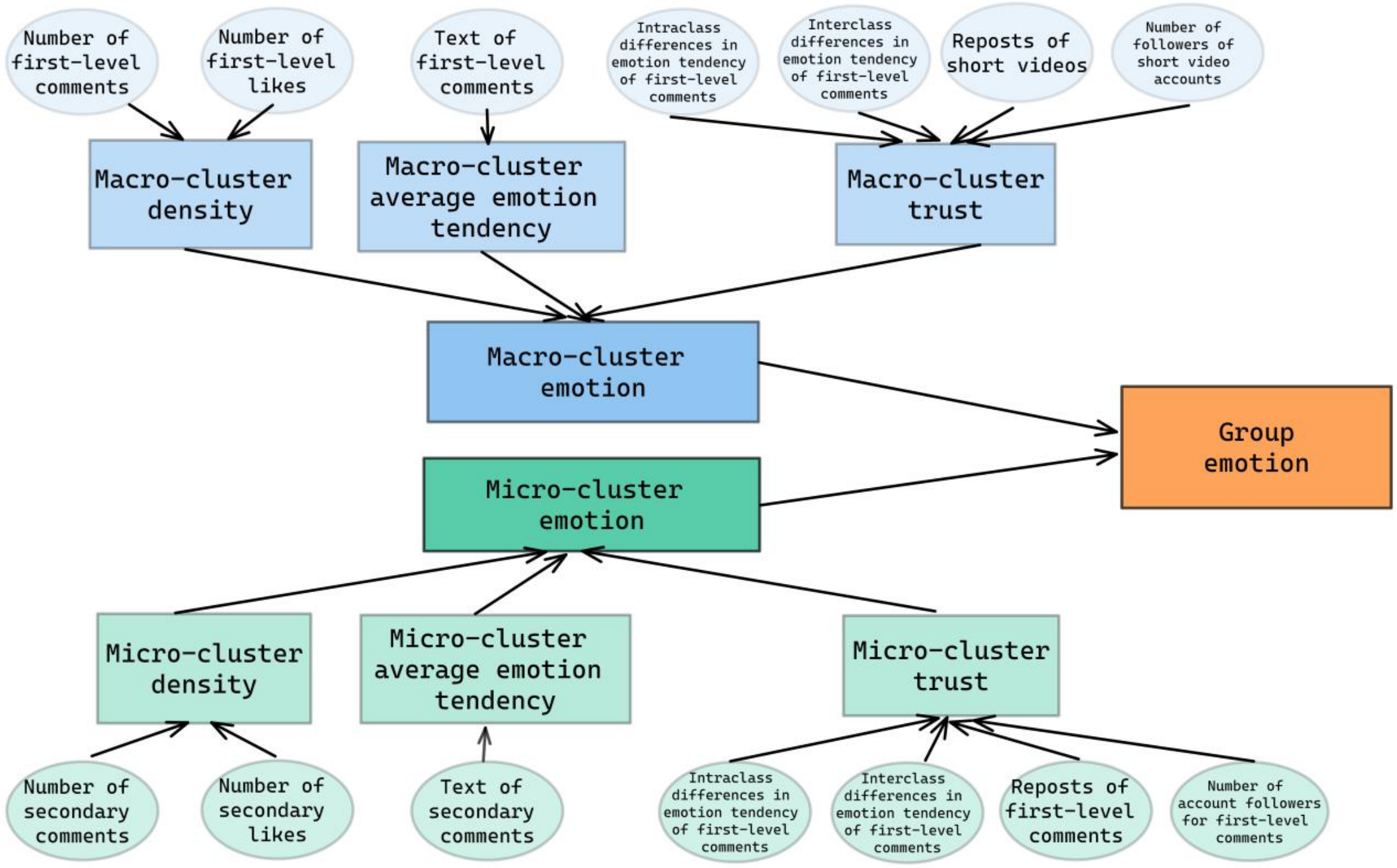

Figure Schematic diagram of the group emotion quantification process in posts

### 3.3. Goup dynamics indicators

UOne of the basic types of classifications for internet users comes from Prensky .It proposed the concepts of digital natives and digital migrants (Prensky, 2001; Prensky, 2006), which divides internet users into two categories based on their characteristics. Specifically, digital natives refer to the younger generation born in the digital age, who often lead the trend of the new media era. In contrast, digital migrants refer to the older age group who have not grown up with electronic tools but gradually become familiar with and adapt to the internet environment during the digitization trend. Therefore, we consider user age as one of the possible dynamic factors that influence group emotion. In addition, since the selected AIGC related keywords include deep learning algorithms and belong to the knowledge technology category, we also consider user education level as a possible dynamic factor that affects group emotion (Kirschner & Karpinski, 2010).

### 3.3.2. Platform Mechanism

Among the three social media platforms of different types in this paper, Douyin and Bilibili are similar in video-sharing. The difference between them is that the length of videos uploaded by users is mainly short or long. In contrast, Weibo is a blog platform that focuses primarily on user socialization, integrated texts, videos and many other media forms, with apparent differences in feedback mechanism and recommendation algorithm design compared to Douyin and Bilibili. For example, the "repost" function on blog platforms and video-sharing platforms represents different meanings. On Weibo, it is used for sharing posts within the platform and is visible to other users. Users can express support or opposition to the reposted content, and this visible reposted content will influence the group opinion formed by the original post content. In contrast, on Bilibili and Douyin, the "repost" function is used for sharing posts to the outside of platform and is not visible. User reposting behavior will not affect the group opinion formed by the original post content, but instead focuses on improving the popularity of the original post content. And unlike Douyin and Bilibili, Weibo's user homepage does not have a recommendation algorithm. All posts on the Weibo homepage come from content published by other users that the user follows, and the platform does not automatically push content that the user has not followed. This can result in an echo chamber where publishers with more followers are more likely to spread post content, while those with fewer followers are less likely to spread post content. In contrast, Bilibili and Douyin will recommend the content that users may be interested in at any time according to their past preferences, which allows accounts with fewer followers to spread posts widely as long as they enter the platform's recommendation system.

Due to the differences in mechanism design on these platforms, we adopt different weight calculation elements when computing group emotions. Additionally, the mechanism design also presents different communication characteristics of the platforms, which is exactly what this paper hopes to obtain from the comparison of different platforms. Previous studies have indicated that emotional content is more likely to spread on the Internet (Berger & Milkman, 2012), but we note that Del, Cinelli has shown that group emotional polarization is more likely to occur on social media platforms like Facebook, which are similar to Weibo (Del Vicario et al., 2016; Cinelli et al., 2021). To verify that differences in platform mechanisms lead to differences in group emotion indeed, we develop two group dynamic indicators, namely the proportion of posts without comments and the proportion of posts from celebrity publishers, as evaluation criteria.

To be specific, the proportion of posts without comments is defined as the ratio of the number of posts without comments to the number of all posts in the dataset for a given topic. For the same topic, the fewer comments on different platforms, the more difficult it is to spread and form a group. This indicator can reflect whether the platform mechanism design can ensure that the content posted by each user is effectively disseminated. The proportion of posts from celebrity publishers is defined as the ratio of the number of posts from celebrity publishers to the number of all posts form a group. Platform certification is a universal mechanism established by all three websites, and verified celebrity publishers are classified into several categories, such as official, commercial, and famous accounts, known as "big v", who are believed to possess higher authority, more fan recognition, and wider spread. The existence of opinion leaders plays a critical role in guiding public opinion on the platform (Wang et al., 2020; Park, 2013; Tsang & Rojas, 2020; Chen et al., 2022), and users are more intend to  follow experts, celebrities, and public figures in their respective fields (Zhang & Pentina, 2012; Marwick, 2013). This indicator can reflect the difference in user influence on different platforms.

## 4. Findings

This section mainly presents the achievements of our work in two aspects: computation results on the dataset, investigations and statistical outcomes of various dynamic indicators.

### *4.1. Different Results of Emotion Calculation in Defferent Websites*

#### *4.1.1.Emotion Values are Similar in Topics,but Different in Websites*

Using the group emotion calculation system in Section 3.2, we first calculated the emotion values of every post group on each platform. The specific total emotion values under each topic are shown in Table 1.

**Table 1**

Group emotion value of each topic

| Topic | Weibo | Bilibili | Douyin |
|---|---|---|---|
| Popular functions (AI painting) | 0.0014 | 0.1218 | -0.0304 |
| Popular functions (AI music) | 0.0671 | 0.2334 | 0.0433 |
| Popular functions (AI writing) | 0.0053 | 0.1431 | -0.0511 |
| Popular functions (AI translation) | 0.0045 | 0.1234 | 0.0324 |
| Popular products (ChatGPT) | 0.0048 | 0.0740 | -0.0871 |
| Popular products (ERNIE Bot) | 0.0111 | 0.0776 | -0.0319 |
| Popular products (New Bing) | 0.0864 | 0.0799 | -0.1826 |
| Popular products (Dreamilyapp) | 0.0049 | 0.1204 | -0.0403 |
| Popular products (Copilot) | 0.0561 | 0.0670 | -0.0293 |
| **Average** | **0.0268** | **0.1156** | **-0.0419** |

The results above indicate that the total group emotion values of Douyin platform towards several emerging technologies are significantly lower than the other two platforms, making it the only platform with negative group emotion values. We can conclude that Douyin shows a negative group emotion towards emerging AI technologies. Compared with Weibo, Bilibili reports generally higher group emotion values, except for a few special topics (New Bing). It can be concluded that the attitude towards emerging technologies among the three experiment platforms is ranked from negative to positive in the order of Douyin, Weibo, and Bilibili. Douyin is a platform that clearly rejects new technologies, while Weibo and Bilibili have an total positive attitude towards new technologies.

To support the conclusions above, we count the amount of positive and negative posts and show the ratio in Table 2. Although the result of group emotion value is a continuous value within the range of [-1,1], we converted this calculation problem into a binary classification problem ignoring the specific size of the emotion value, and defined posts with a group emotion value greater than 0 as positive emotion posts and posts with a group emotion value less than or equal to 0 as negative emotion posts, in order to display a platform's positive or negative attitude towards the topic more directly.

**Table 2**

the proportional chart of positive and negative posts

| Topic | Weibo | Bilibili | Douyin |
|---|---|---|---|
| Popular functions (AI painting) | 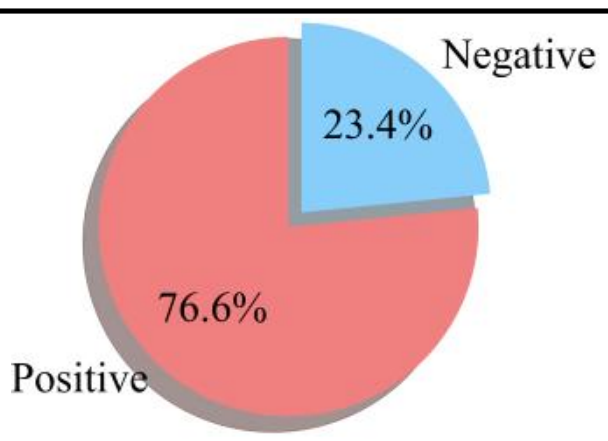 | 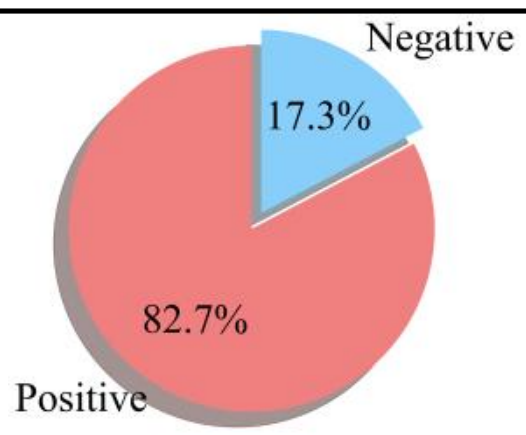 | 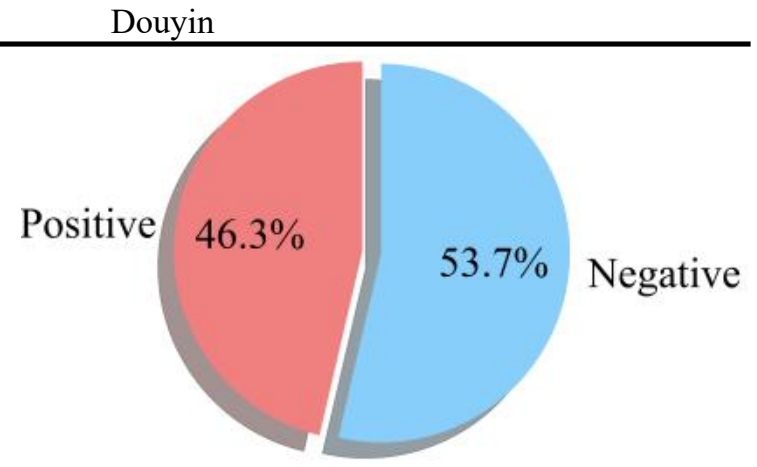 |
| Popular functions (AI music) | 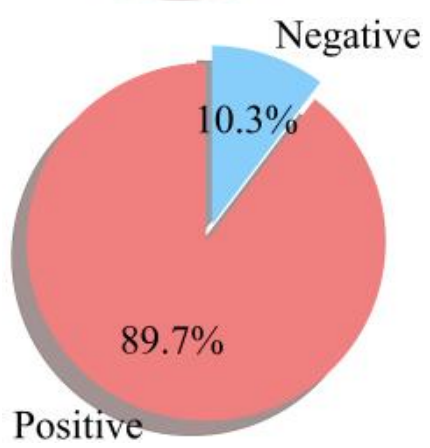 | 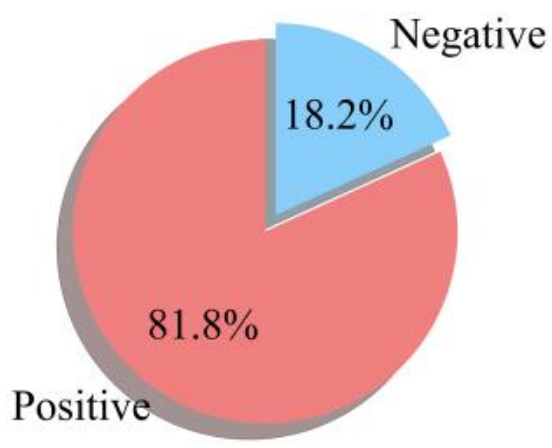 | 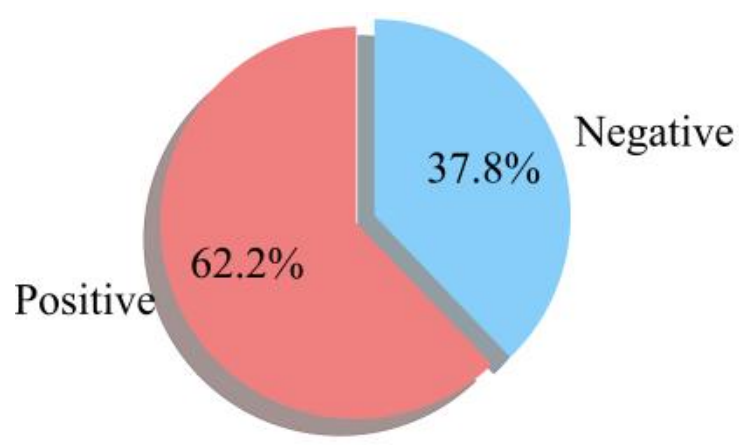 |

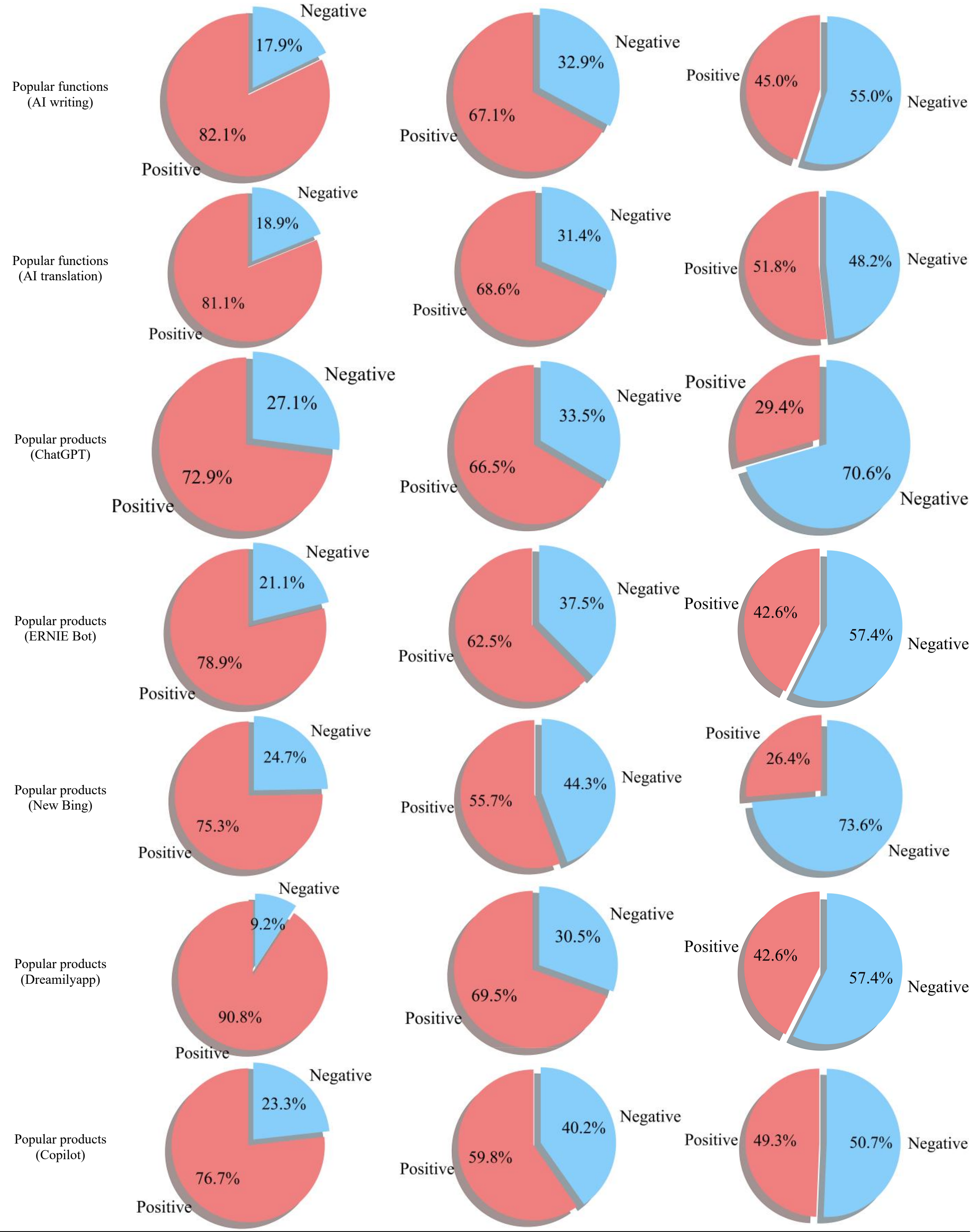

In Table 2, it can be observed that there are emotion differences between experiment platforms that are not affected by topic differences: Douyin records the highest proportion of negative posts across all topics, excepting the the topic of AI art (music), with more than half of negative posts which is significantly higher than the other platforms. Douyin is the only platform with a larger number of negative than positive emotion posts. On Weibo and Bilibili, the proportion of positive posts exceeds 50% on all topics, these two platforms still hold optimistic views towards emerging AI technologies, while the proportion of negative posts on Bilibili is higher than that on Weibo. It is notable that although Bilibili's tendency towards a higher proportion of negative emotion posts on Weibo is largely unaffected by the topic difference, only under the topic of AI art (painting), the proportion of negative posts on Weibo is much higher than that on Bilibili. That is consistent with the platform difference shown by the group emotion values of each topic.

The total group emotion value of a topic is the average emotion value of all posts under that topic, which is influenced by the ratio of positive to negative emotion posts. We made a polarity judgment on continuous values, ignoring the magnitude of group emotion values, as well as the impact of posts with absolute values close to 0, i.e., neutral posts for the ratio. We infer that a higher proportion of neutral emotion posts on Bilibili leads to a higher negative post ratio compared to Weibo, while the topic emotion values are higher on Bilibili than on Weibo.

Apart from the platform differences, we can also summarize the current public opinion on AI-related emerging technologies from the differences in indicators across various topics. Based on the amount of data that can be crawled, ChatGPT and AI arts of painting are the two most popular discussion topics, followed by ERNIE Bot, with a large number of participants and controversies. The public attitude towards AI conversation products including ChatGPT and ERNIE Bot is more negative than towards AI arts including AI painting and AI music. Compared to Douyin, the emotion values of Weibo and Bilibili are closer to each other, and some topics have eased the differences between platforms to some extent. For example, Weibo and Bilibili show a higher emotion value trend for AI music compared to AI painting and writing, while for ChatGPT, it is lower on both platforms. We can explore the common emotion differences of internet users towards different AI products from these findings. It is particularly notable that Bilibili shows similar group emotion values for the three different topics under the AI conversation category, while Weibo shows significantly different group emotion tendencies for the three topics under the AI conversation scope. Further research is needed to investigate the group dynamic factors that may cause such topic differences.

### *4.1.2. Group Emotion Value Polarize with Cluster Size in Weibo*

In the computation of group emotion, cluster density is of great significance as it reflects not only the redundancy of the group, but also the size of each post cluster, indicating the popularity and spread of a post. Cluster density is calculated as the sum of replies and likes, with likes offering an intuitive measure of the popularity of a single post. As the proposed group emotion calculation system is based on post comments, cluster density is equivalent to the cluster size of each primary post, with larger clusters tending to be more widely spread. Due to the features of internet platforms, there is a large difference in cluster size between different posts in the data obtained in this experiment, which can be several thousand times.

The relationship between cluster size and group emotion is illustrated in Table x. To investigate their relationship, we arranged content clusters on the horizontal axis from left to right based on their group emotion values. The horizontal axis represents the ranked cluster number after sorting by group emotion value, and the vertical axis represents the normalized cluster size within the range of [0,1]. The color of the column represents group emotion, with red indicating positive emotion, blue indicating negative emotion, and darker colors indicating larger absolute values. Column close to white can be judged as a cluster holding neutral emotion. Due to significant differences in cluster size, some very small clusters were omitted from the plot, and some small clusters were not clearly represented. Results are presented for only the ChatGPT, New Bing, and AI Writing topics due to space limitations, with results for other topics included in Appendix A.

**Table x**

Relationship between group emotion value and post popularity

| Topic | Weibo | Bilibili | Douyin |
|---|---|---|---|
| Popular functions (AI writing) | 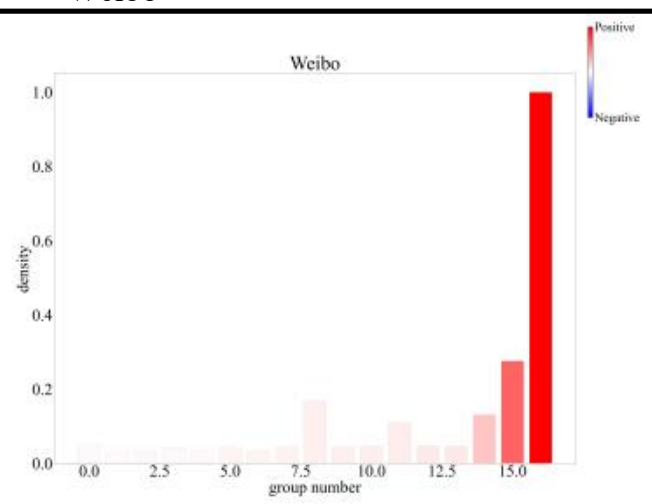  | 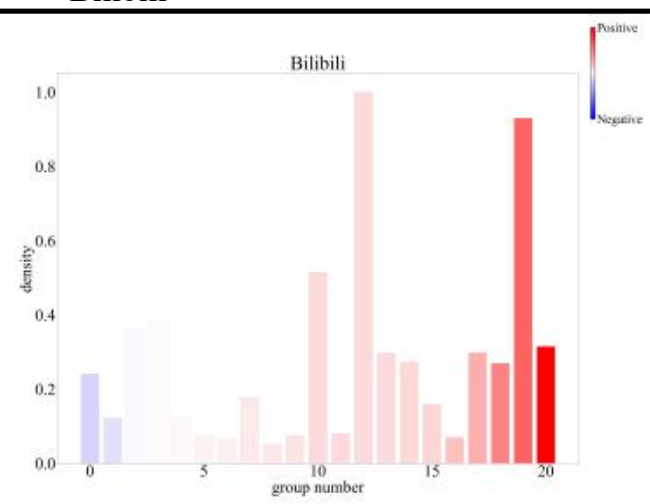  | 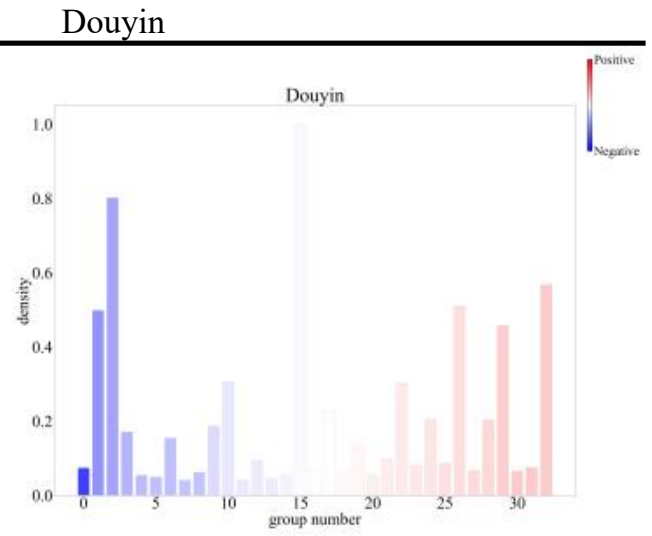  |
| Popular products (ChatGPT) | 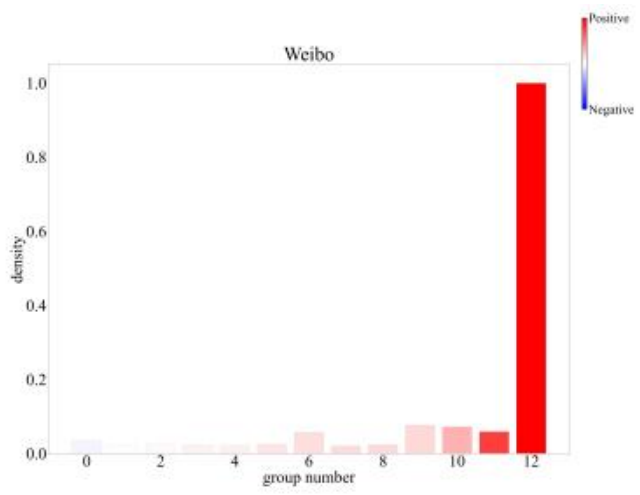  | 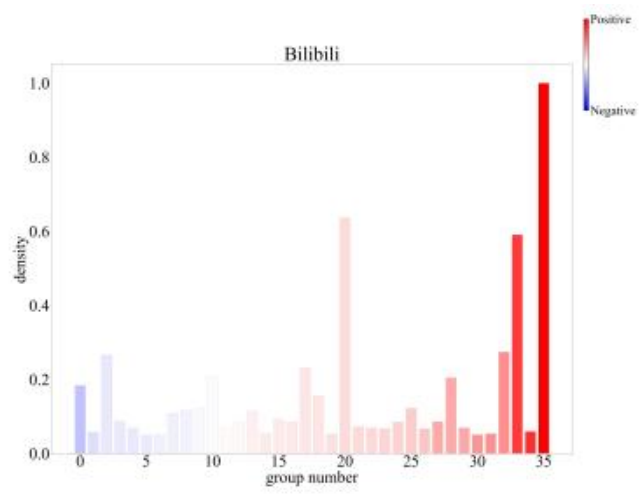  | 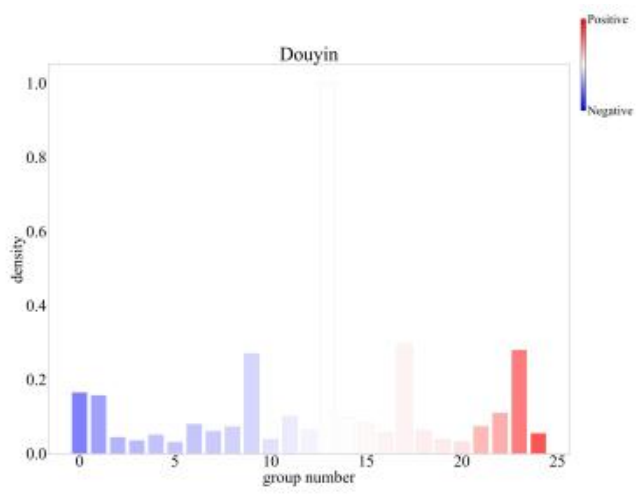  |
| Popular products (New Bing) | 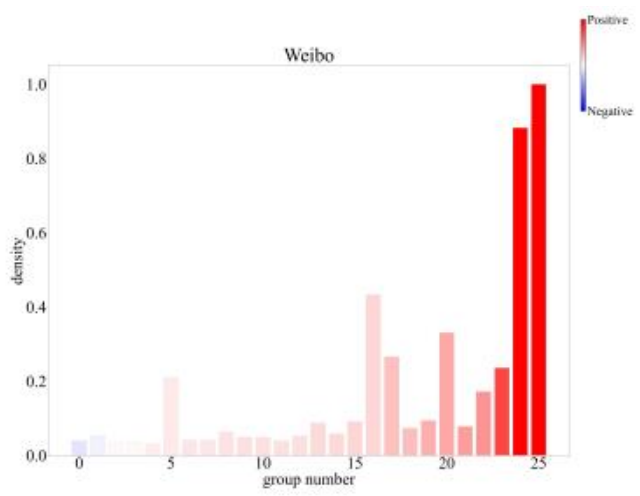  | 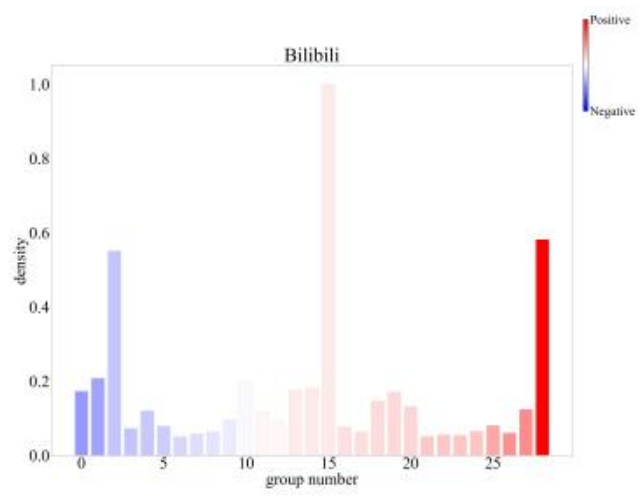  | 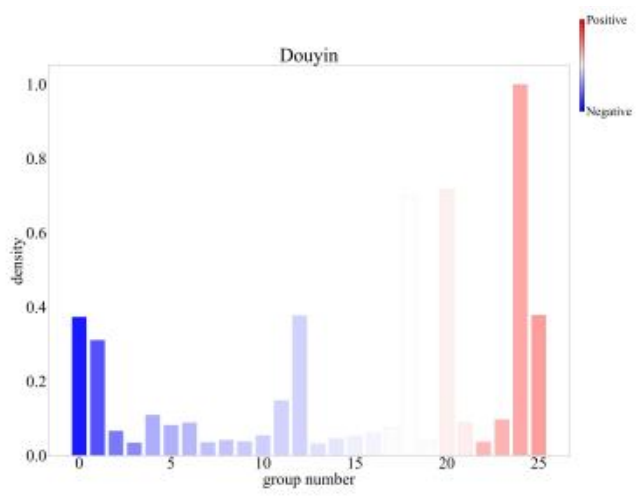  |

The figure shows that there are significant differences in the distribution characteristics of cluster popularity on various platforms. Weibo generally displays unipolar or bipolar group emotion distributions in all topics, with popular clusters usually having a bigger emotion absolute values, and there are no popular clusters with a neutral emotion tendency. In contrast, the cluster popularity of Bilibili and Douyin is more evenly distributed with the emotion value of the cluster and the high-emotion value clusters are not necessarily more popular. There are clusters in super-large sizes with an emotion tendency towards neutrality, unlike Weibo, which exhibits a clear polarizing tendency in all topics. Weibo exhibits a clear polarizing tendency in all topics, while group polarization also occurs in Bilibili in some topics, but the distribution of clusters in Bilibili is often more even than Weibo and the trend of cluster size is not clear under the same topic. This suggests that extreme emotions are more likely to be welcomed and popular on Weibo, and posts that can trigger large-scale spread on Weibo often exhibit stronger and more extreme group emotion tendencies. But posts with a neutral or indistinct group emotion tendency can also generate broader discussions and higher popularity on Bilibili and Douyin.

**Table x**

Skewness coefficient of each topic

| Topic | Weibo | Bilibili | Douyin |
|---|---|---|---|
| Popular functions (AI painting) | -3.6079 | 1.4591 | 1.2303 |
| Popular functions (AI music) | 3.6714 | -0.4498 | -0.3115 |
| Popular functions (AI writing) | 18.2295 | -0.1954 | -0.3619 |
| Popular functions (AI translation) | 7.6222 | 0.0679 | 0.3442 |
| Popular products (ChatGPT) | 18.9778 | 0.6070 | 0.1073 |
| Popular products (ERNIE Bot) | 9.7365 | -0.1625 | -1.9230 |
| Popular products (New Bing) | 3.3969 | 0.0897 | 0.2407 |
| Popular products (Dreamilyapp) | 24.7990 | -0.1173 | 0.0786 |
| Popular products (Copilot) | 4.9084 | 0.7125 | -0.1001 |
| **MAE** | **10.5450** | **0.4290** | **0.5220** |

## *4.2. Similar Performance between Group Dynamics Indicators and Group Emotions Within a Same Website*

### *4.2.1. User Education Level*

Figure 6 displays and compares the education level distribution data of users in the three platforms, and from this analysis, important patterns can be summarized. The education level composition of users on Weibo and Bilibili is relatively similar: more than half of the users possess a bachelor's degree or higher, signifying an majority, while users with junior high school and high school education levels account for a similar proportion. Bilibili possesses a significantly higher proportion of users with a bachelor's degree or higher, reaching 80%, which is markedly higher than that of Weibo, highlighting the significant characteristic of high education level among Bilibili users. In contrast, the education background composition of Douyin users is quite different: only 28.6% of users hold a bachelor's degree or higher, and the proportion of users with junior high school and high school education levels is similar, with the highest proportion of users being of high school education level. Users with a high school education level or lower account for more than half, and comprise the majority.

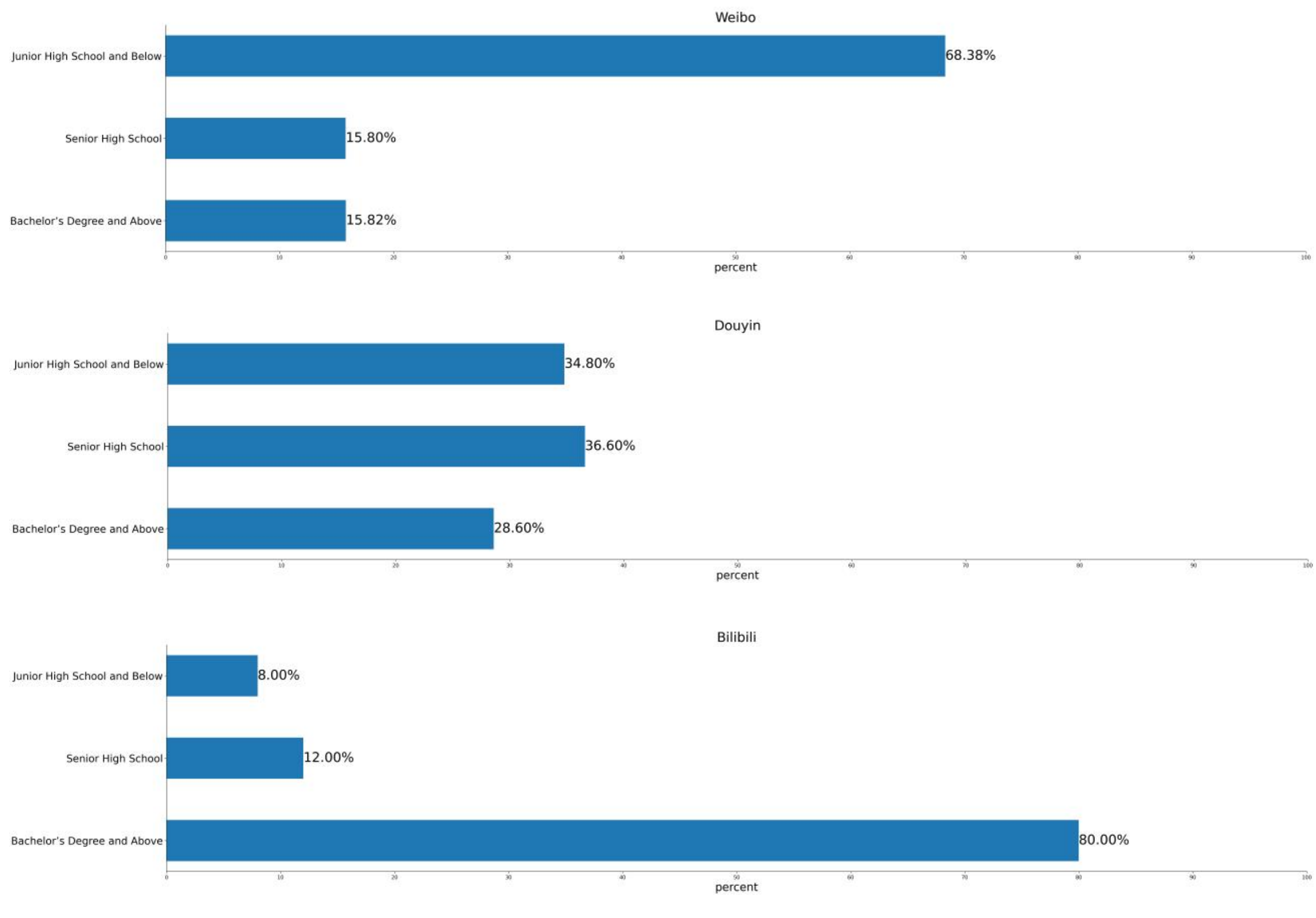

**Fig. 6.** Proportion of users' education level on three websites.

### *4.2.2. User Age Group*

Fig. 7. compares and contrasts the proportion data of users across different age groups on the three platforms, among which Bilibili shows a significant difference from the other two platforms: users from the youth group of 25 years old and below form a majority, reaching as high as 74.6% in Bilibili, and the proportion of middle-aged users aged 36 years and above is only half that of Weibo and one-third that of Douyin. Bilibili's prevailing users are predominantly the Z generation characterized by youth and high education, many of whom were born between 1993 and 2005 and possess the characteristics of digital natives (Turner, 2015; Cilliers, 2017). This is a significant feature that distinguishes Bilibili from other platforms. In contrast, the age group composition of users on Douyin and Weibo is more evenly and similar.

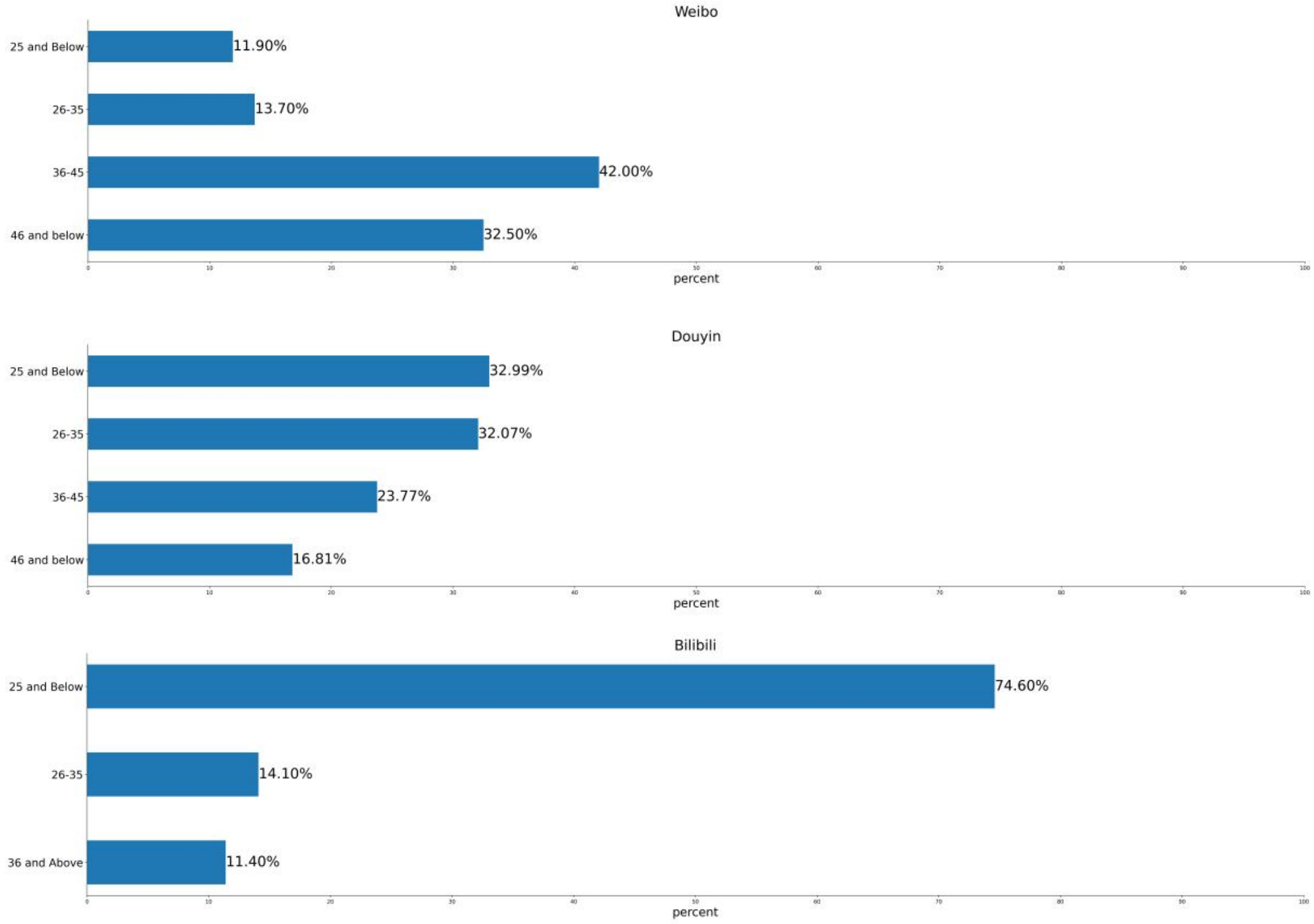

**Fig. 7.** Proportion of users' age group on three websites.

*4.2.3.Proportion of Posts without Comments*

According to the instructions in section 3.3.2, we calculated the proportion of posts without comments for each platform in the dataset, and the results are shown in Table 3. From the table, we can observe that the proportion of posts without comments on Weibo is all above 70%, and even reaching as high as 95.5%, implying that posts that cannot achieve high popularity and effective spread occupy the majority of Weibo's content production, making it difficult for ordinary users to form post clusters on Weibo. Compared with Weibo, the proportion of posts without comments on Douyin is relatively low, remaining at around 30%, and the proportion of posts without comments on Bilibili is greatly affected by topic popularity, but is still lower than that of Weibo.

**Table 3**

Proportion of posts without comments of each topic

| Topic | Weibo | Bilibili | Douyin |
|---|---|---|---|
| Popular functions (AI painting) | 78.1% | 26.9% | 28.1% |
| Popular functions (AI music) | 81.3% | 65.0% | 9.0% |
| Popular functions (AI writing) | 78.6% | 68.0% | 13.8% |
| Popular functions (AI translation) | 82.6% | 20.8% | 11.1% |
| Popular products (ChatGPT) | 77.1% | 32.5% | 16.7% |
| Popular products (ERNIE Bot) | 76.7% | 41.4% | 12.8% |
| Popular products (New Bing) | 95.8% | 22.7% | 25.7% |
| Popular products (Dreamilyapp) | 75.3% | 61.6% | 32.4% |
| Popular products (Copilot) | 95.5% | 33.9% | 23.1% |
| **Average** | **82.33%** | **41.42%** | **19.19%** |

*4.2.4.Proportion of Celebrity Publishers*

We have calculated the ratio of celebrity publishers to ordinary users for each platform in the dataset, as shown in Table 4. Among the posts forming a cluster we crawled, the percentage of celebrity publishers on Weibo far exceeds the other two platforms, accounting for more than half of all users, which is significantly different from the other two platforms. And Bilibili's celebrity publishers ratio is the lowest among experiment platforms.

**Table 4**

Proportion of posts published by celebrity users of each topic

| Topic | Weibo | Bilibili | Douyin |
|---|---|---|---|
| Popular functions (AI painting) | 56.9% | 19.0% | 22.0% |
| Popular functions (AI music) | 58.8% | 8.8% | 51.4% |
| Popular functions (AI writing) | 52.8% | 11.8% | 25.4% |
| Popular functions (AI translation) | 62.3% | 10.9% | 34.8% |
| Popular products (ChatGPT) | 52.8% | 11.8% | 15.3% |
| Popular products (ERNIE Bot) | 81.6% | 10.3% | 40.5% |
| Popular products (New Bing) | 82.4% | 0.7% | 10.9% |
| Popular products (Dreamilyapp) | 62.2% | 3.4% | 2.7% |
| Popular products (Copilot) | 88.4% | 13.4% | 18.0% |
| **Average** | **66.47%** | **9.9%** | **24.56%** |

## 5. Discussion

Based on the experimental results in Section 4, we have summarized several conclusions. In this section, we apply these conclusions to infer the corresponding group emotions and group dynamics indicators.

### *5.1. Summary of Findings*

The results presented in Section 4.1.1 demonstrate that based on the data, the group emotion value of Douyin is the lowest among all the three platforms and exhibits negative attitudes, while the group emotions on Weibo and Bilibili display a consistent and positive trend. That is consistent with the results shown in Section 4.2.1: the education level of all users on Douyin is generally low, while the education background structure of all users on Weibo and Bilibili is similar. It effectively suggests that the education level of users is a critical dynamic factor in the formation of group emotion for AIGC topics. And the data in Section 4.2.2 shows a significant difference in the age group of users on Bilibili compared to Weibo and Douyin, which is also corresponds to the conclusion drawn in Section 4.1.1 that the group emotion value on Bilibili is the highest among the three platforms. The proportion of teenagers on Bilibili is much higher than the other two platforms, indicating that the group in this age tends to appear a more positive, tolerant, and rational attitude towards emerging technology topics. The age of users is one of the dynamic factors that affect the formation of group emotion.

The data presented in Section 4.1.2 reveals a distinctive feature of Weibo that clearly sets it apart from the other two platforms: Weibo is more susceptible to group emotion polarization. By comparing these findings with platform mechanism results in Sections 4.2.3 and 4.2.4, it becomes apparent that Weibo also displays features that are different from the other two platforms: the proportions of posts without comments and celebrity publishers are both extremely high. This suggests that the platform mechanism design of Weibo differentiates the impact scope of different users, and ordinary users' post content cannot be effectively spread. Posts by ordinary users that can trigger wide dissemination phenomena are more likely to contain controversial attitude. Such platform mechanisms constitute the main dynamic factor for the spread of extreme emotion on Weibo.

### 5.2. Implications

First in theory, we have confirmed that education level of website users is the primary dynamic factor behind the formation of corresponding group emotions for emerging scientific and technological topics such as AIGC, while platform mechanism design can lead to distinctive dissemination characteristics on the websites. As demonstrated in this paper, the platform mechanism design makes it easier for extreme emotion to spread on Weibo. In the context of the popularity of new AIGC technologies, the research findings of this paper have significant practical implications for public opinion regulation and control, as well as cross-platform information dissemination. For the industry impact and social issues brought about by AI technology, the public needs to continuously improve their media literacy, and rationally analyze and judge information. Media platforms need to strengthen their self-governance, and build a good and equal discussion environment to reduce the degree of information dissemination bias. The governments need to carry out top-level design for public opinion guidance based on the platform mechanism and user profile, and selectively disseminate information according to platform differences.

## 6. Conclusion

In this paper, based on each individual content of a web post, we establish a virtual group in online space, using user interactive comments as the basic element. And we simulate a two-stage emotion propagation chain from individual small micro-clusters of comments and replies to the whole large macro-cluster of all comments, designing an end-to-end complete social network group emotion calculation system at the same time, which is used to investigate the public opinion status of the popular AIGC technology. Based on the theory of group dynamics, we also design multiple group emotion dynamic quantification indicators. By comparing the indicator performance of three different types of social media on various relevant topics, this paper summarizes the characteristics of three social media platforms in terms of their communication dynamics, exhibiting the public's emotion attitude, discussion intensity, degree of dissent, and different emotion reactions towards different AIGC categories. There are two main shortcomings of this study: 1) Affective computing for comment texts only uses the snownlp package, and the snownlp dictionary is based on the replies of e-commerce platforms, which may result in bias in emotion value calculation. In the future, we plan to introduce more accurate text emotion calculation methods to improve the accuracy of the group emotion model. 2) We do not introduce group dynamics analysis for topics. Due to the length limitation of this paper, we mainly focus on comparing the differences between platforms and do not further analyze the dynamic differences in topics. Subsequent work needs to combine the experiment data of this paper to further analyze the dynamic factors causing topic emotion differences.

Therefore, it is necessary to summarize the public's emotion attitudes towards AIGC on social media and explore the internal causes of emotional formation. In terms of theoretical significance, this study is aimed to provide a new perspective for group dynamics theory research of the formation of group emotions. In terms of application significance, we hope the present work can give decision-making references for public opinion guidance and help offer solutions for possible social problems during the technology transition period.

## CRediT authorship contribution statement

**Qiaolei Jiang:** Conceptualization, Methodology, Investigation, Formal analysis, Supervision, Writing – original draft, Writing – review & editing, Funding acquisition. **Yadi Zhang:** Conceptualization, Data curation, Investigation, Formal analysis, Software, Writing – original draft, Writing – review & editing. **Wenjing Pian:** Investigation, Validation, Resources, Project administration, Writing – review & editing, Funding acquisition.

## Data Availability

Data will be made available on request.

## Acknowledgement

This study was supported by the National Social Science Foundation of China (No. 18ZDA307) and the National Natural Science Foundation of China (No. 71904028).

## Appendix A. Relationship between group emotion value and post popularity on all experiment topics

| Topic | Weibo | Bilibili | Douyin |
| --- | --- | --- | --- |
| Popular functions (AI painting) | 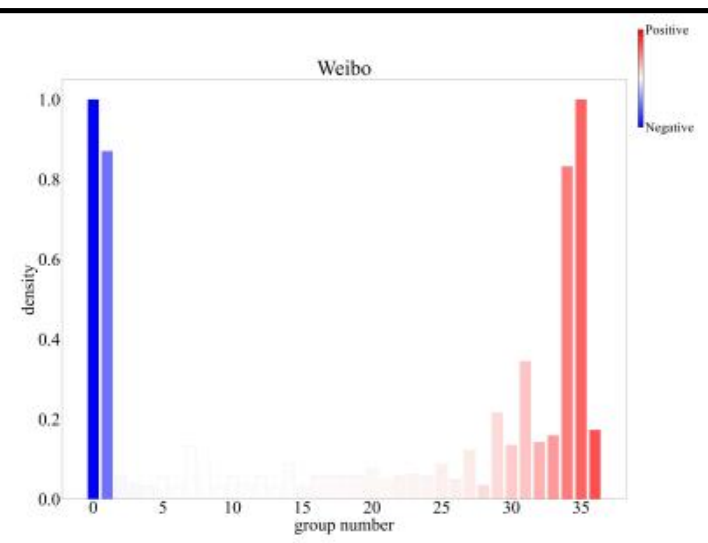 | 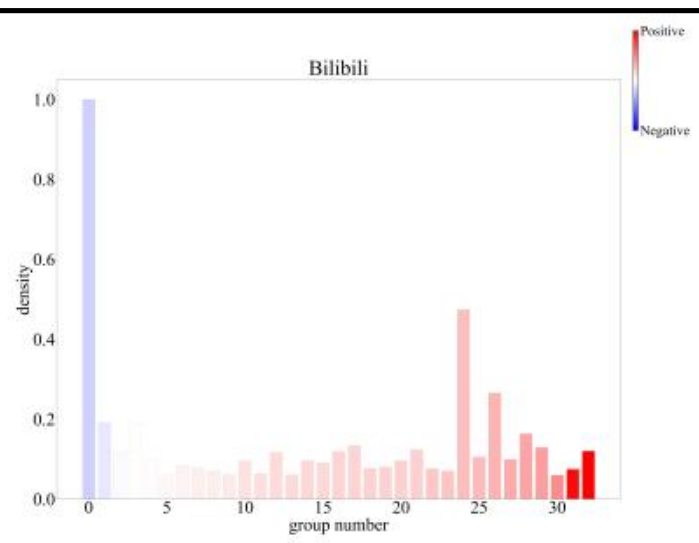 | 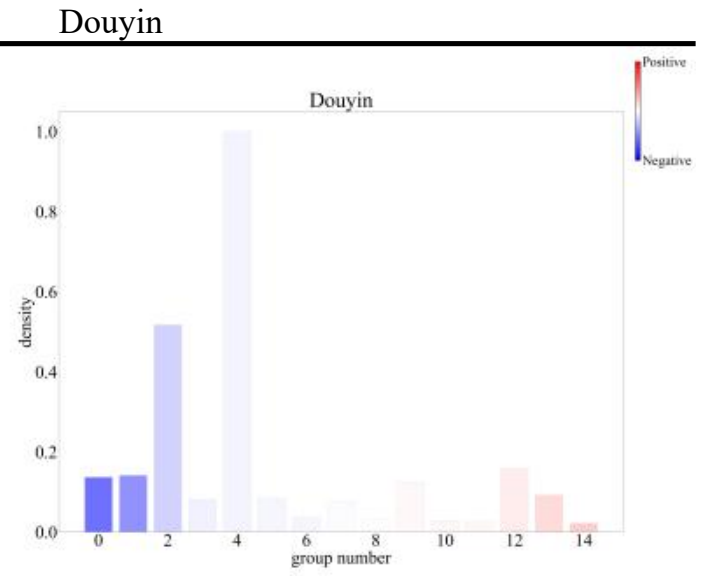 |

(continued on next page)

(*continued* )

| Topic | Weibo | Bilibili | Douyin |
|---|---|---|---|
| Popular functions (AI music) | 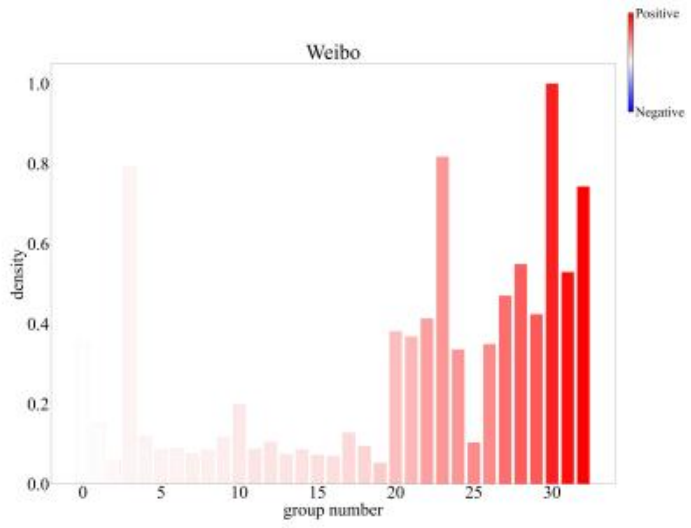 | 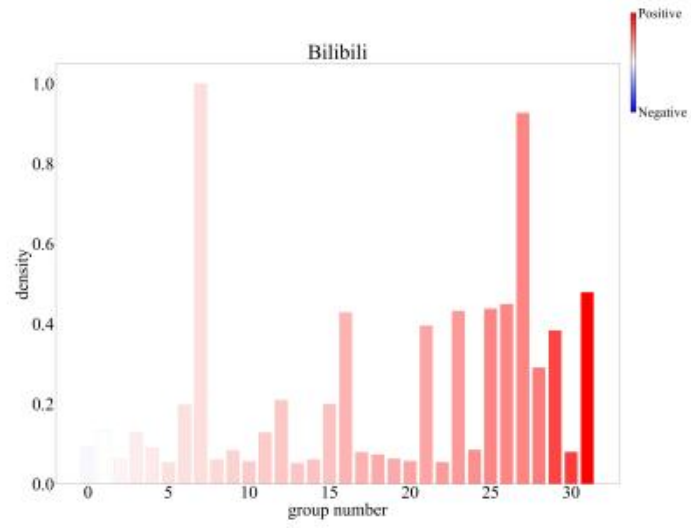 | 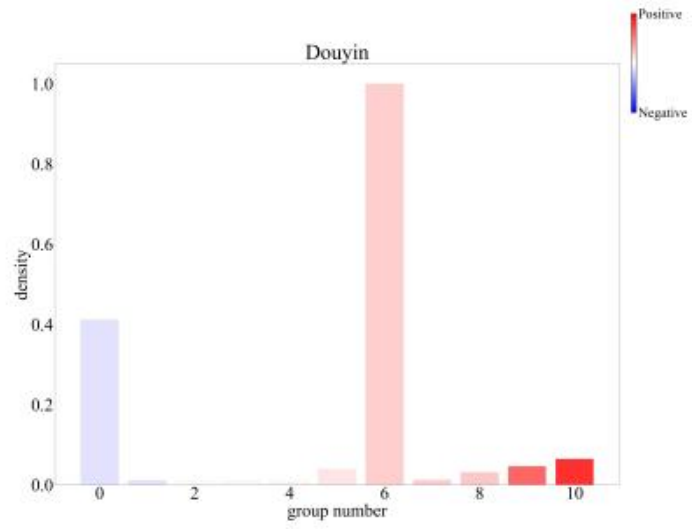 |
| Popular functions (AI writing) | 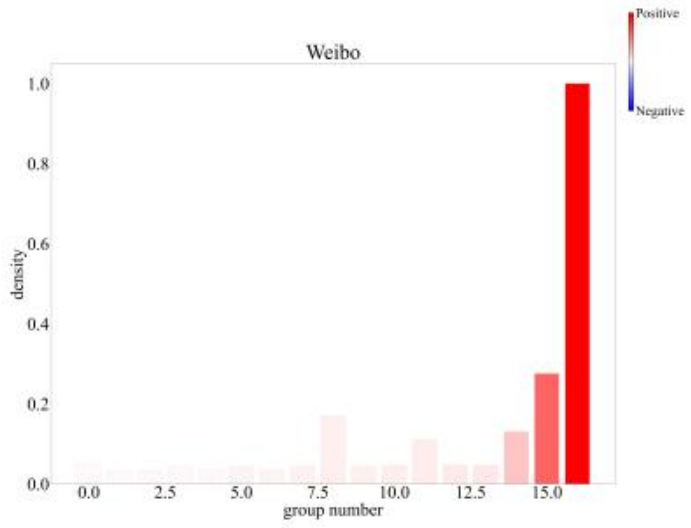 | 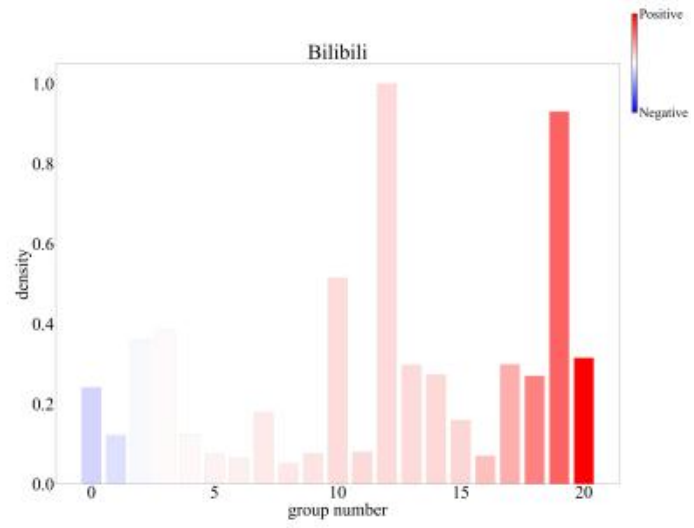 | 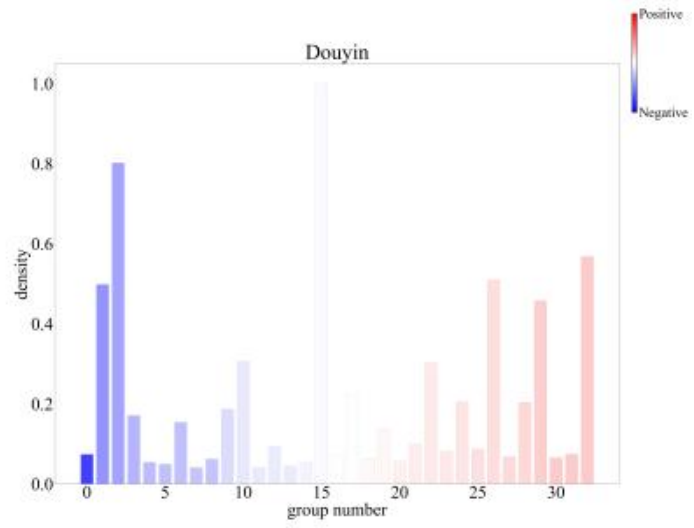 |
| Popular functions (AI translation) | 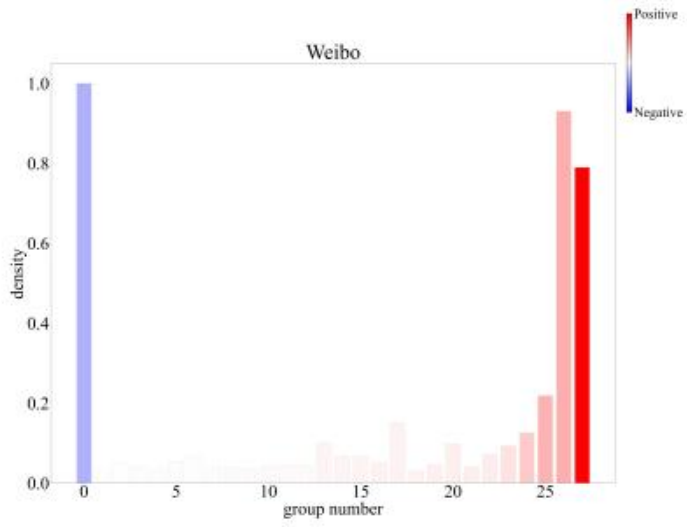 | 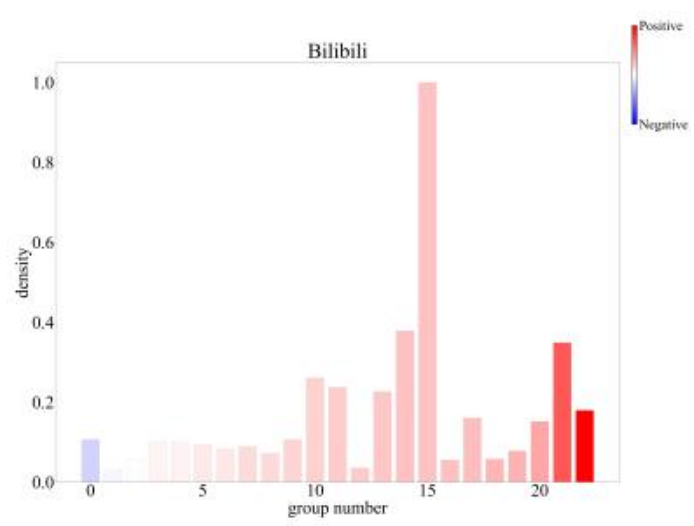 | 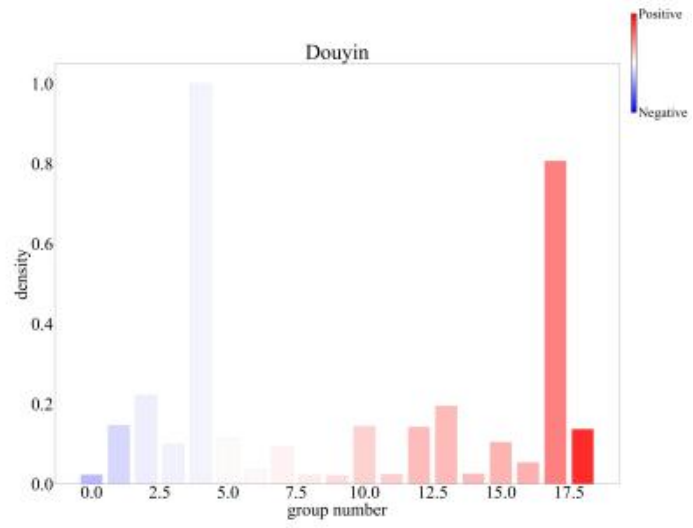 |
| Popular products (ChatGPT) | 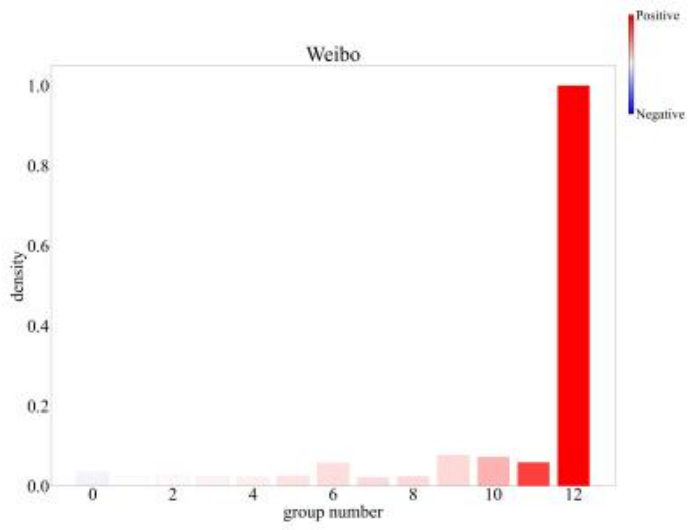 | 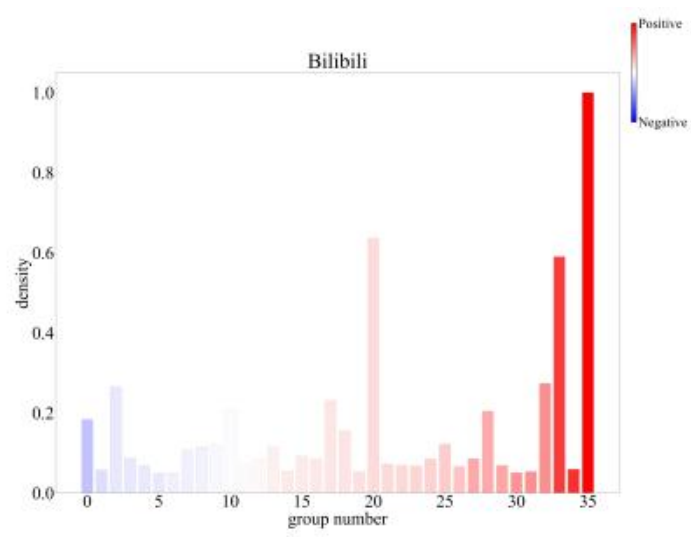 | 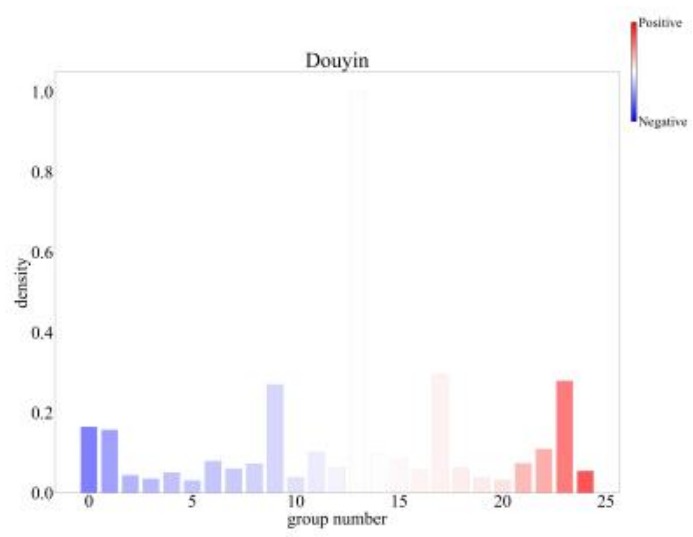 |
| Popular products (ERNIE Bot) | 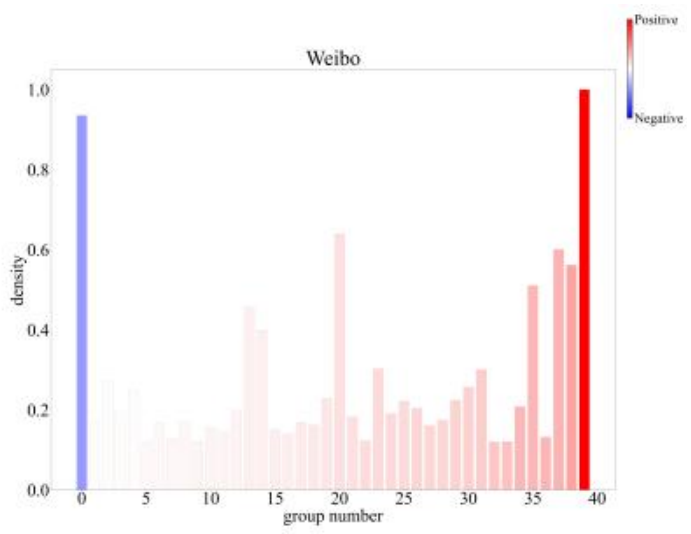 | 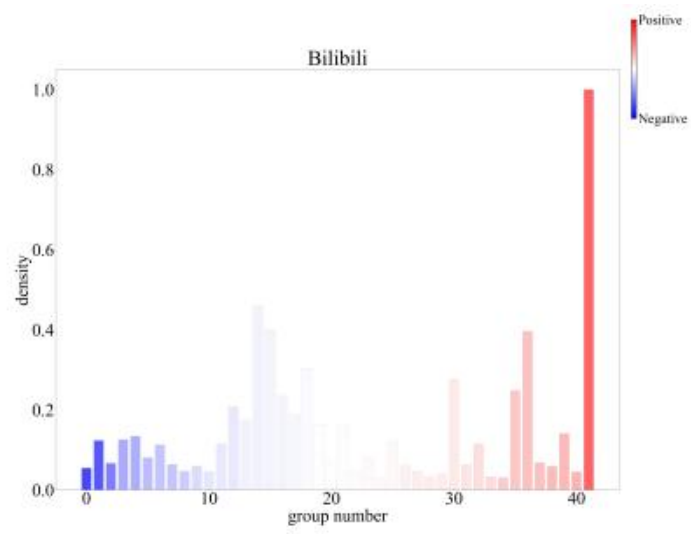 | 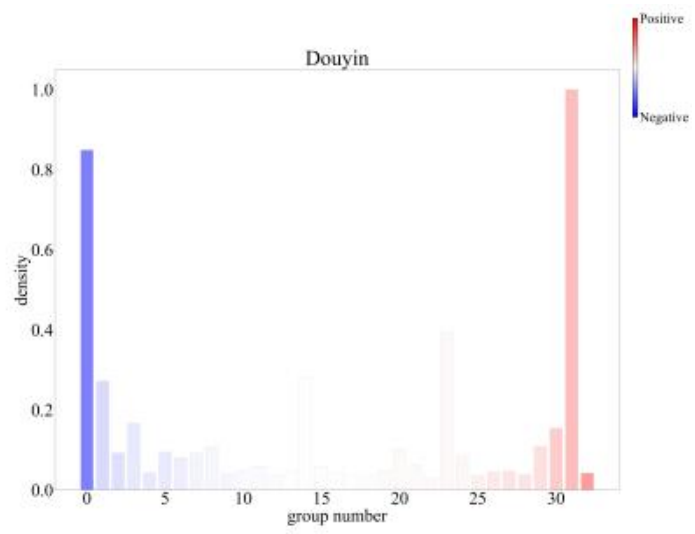 |
| Popular products (New Bing) | 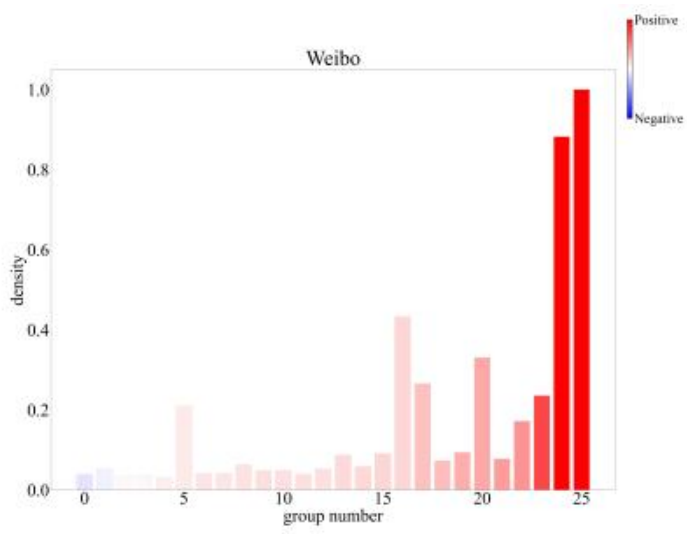 | 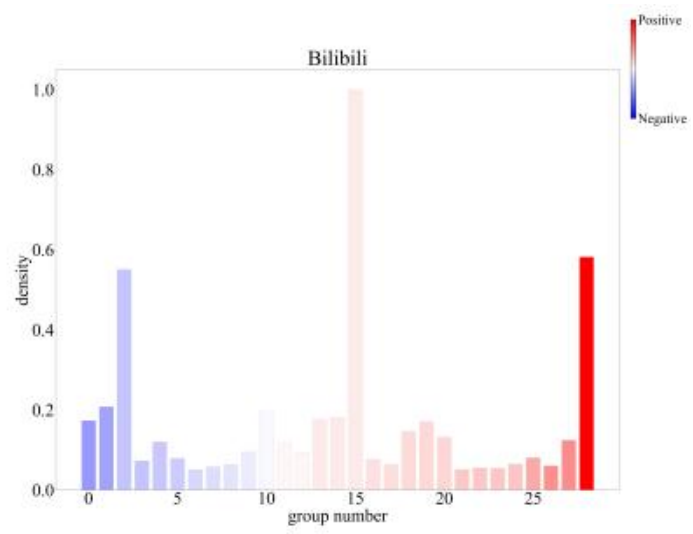 | 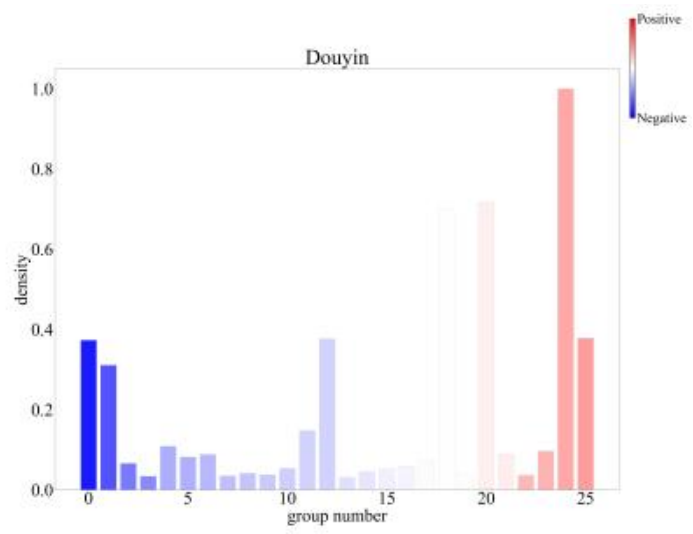 |

(continued on next page)

(*continued* )

| Topic | Weibo | Bilibili | Douyin |
|---|---|---|---|
| Popular products (Dreamilyapp) | 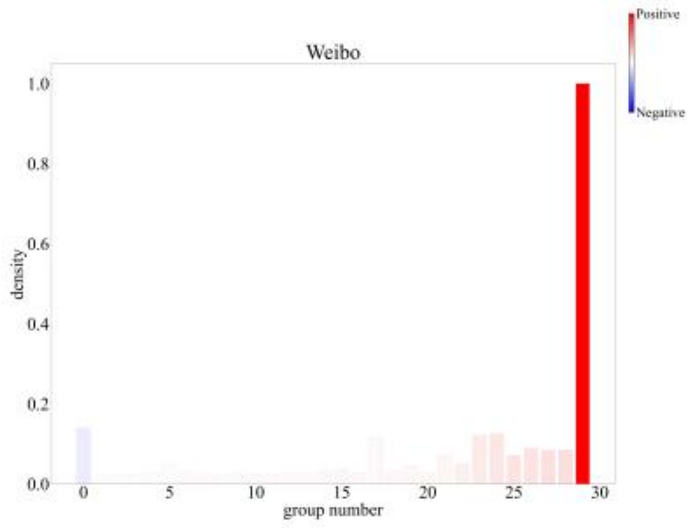 | 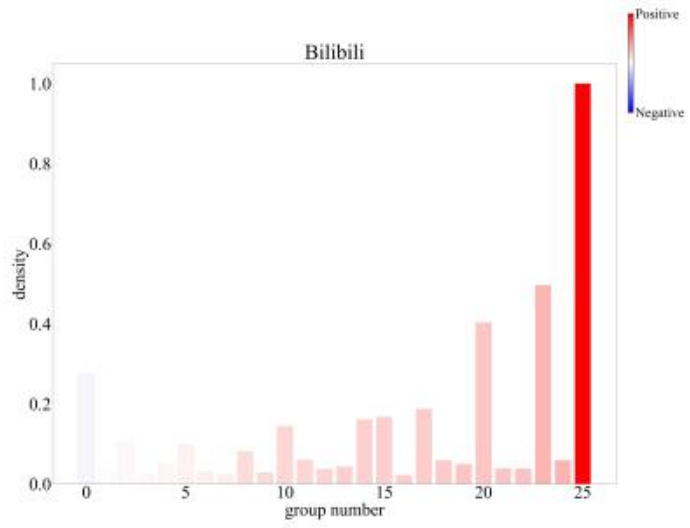 | 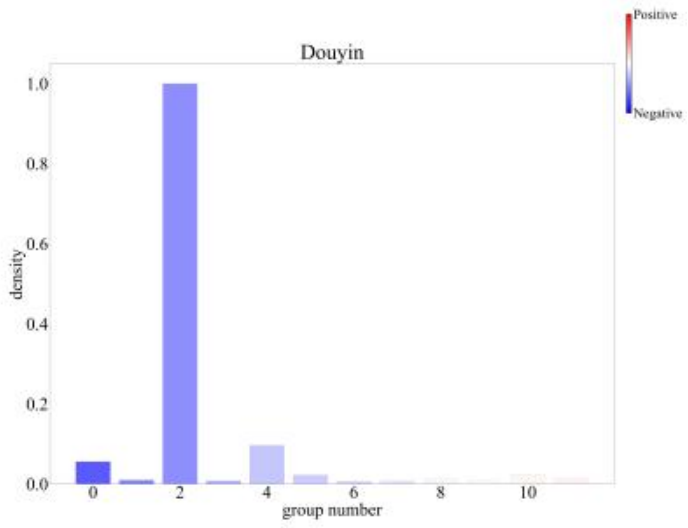 |
| Popular products (Copilot) | 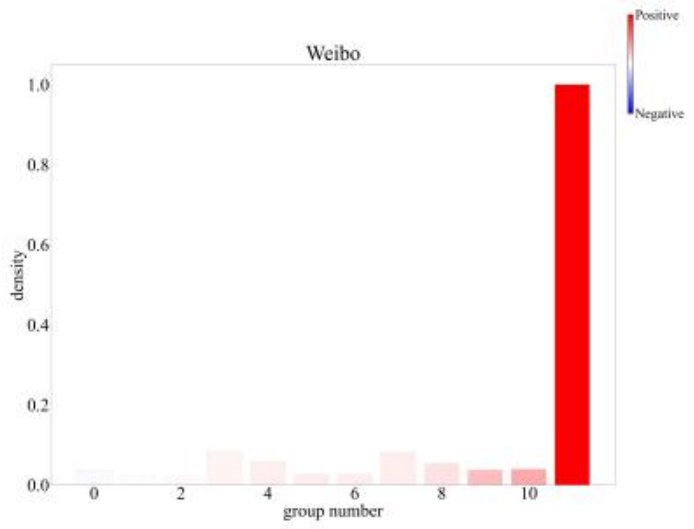 | 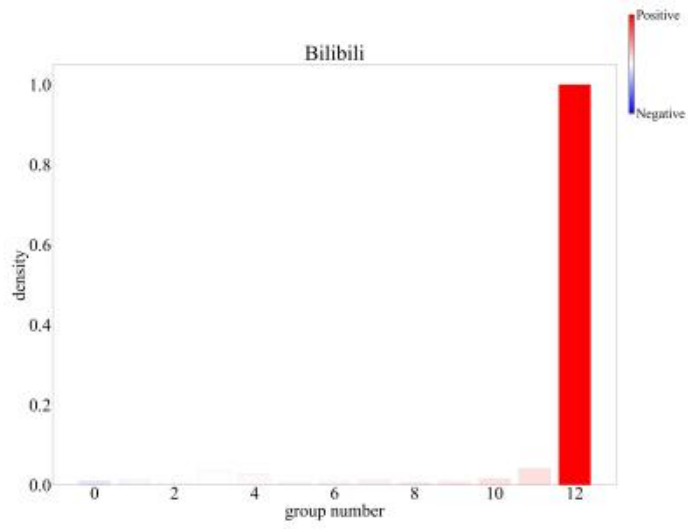 | 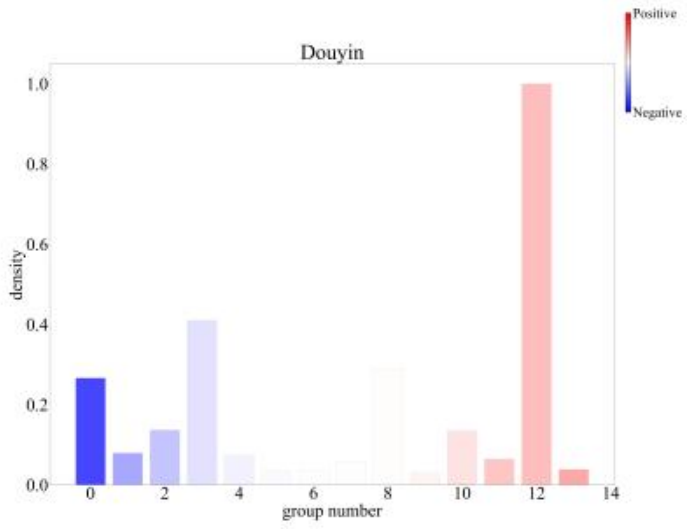 |